\documentclass[aps,prb,10pt,twocolumn]{revtex4-2}

\usepackage{graphicx}
\usepackage{amssymb,amsmath}
\usepackage{color}
\usepackage{braket}
\usepackage{bm}
\usepackage{dcolumn}

\newcommand{\bfr}{{\bf r}}

\newcommand{\bfk}{{\bf k}}

\newcommand{\bfG}{{\bf G}}

\begin{document}

\title{Optical properties of CsCu$_2$X$_3$ (X=Cl, Br and I): A comparative study between hybrid time-dependent density-functional theory and
the Bethe-Salpeter equation}

\author{Jiuyu Sun}
\affiliation{Department of Physics and Astronomy, University of Missouri, Columbia, Missouri 65211, USA}

\author{Carsten A. Ullrich}
\affiliation{Department of Physics and Astronomy, University of Missouri, Columbia, Missouri 65211, USA}

\date{\today }

\begin{abstract}
The cesium copper halides CsCu$_2$X$_3$ (X=Cl, Br and I) are a class of all-inorganic perovskites with interesting and potentially useful optical properties,
characterized by distinct excitonic features. We present a computational study of the optical absorption spectra of CsCu$_2$X$_3$, comparing time-dependent
density-functional theory (TDDFT) and the Bethe-Salpeter equation (BSE), using $GW$ quasiparticle band structures as input.
The TDDFT calculations are carried out using several types of global hybrid exchange-correlation functionals. It is found that an admixture
of nonlocal exchange determined by the dielectric constant produces optical spectra in excellent agreement with the BSE. Thus, hybrid
TDDFT emerges as a promising first-principles approach for excitonic effects in solids.
\end{abstract}

\maketitle

\section{Introduction}

Low-dimensional (quasi-2D, 1D, and 0D) hybrid organic-inorganic lead halides have been intensely investigated as promising luminescent
materials \cite{ref1,ref2,ref3,ref4,ref5,ref6}.
The low dimensionality of these structures gives rise to quantum confinement effects, which yield more stable excitons at room temperature, and therefore enhanced
luminescence properties \cite{ref7,ref8,ref9}. However, the practical applications of these materials are limited by the instability and toxicity
caused by the organic constituents and the presence of Pb ions \cite{Etgar2018,Fu_AdvSci2018}.

Recently, all-inorganic low-dimensional Cu halides have attracted increasing attention, such as quasi-0D Cs$_3$Cu$_2$X$_5$ \cite{Cs3Cu3X5_ref1,Cs3Cu3X5_ref2,Cs3Cu3X5_ref3,Cs3Cu3X5_ref4}, quasi-1D Rb$_2$CuX$_3$ \cite{Rb2CuBr3_ref1,Rb2CuBr3_ref2}, and quasi-1D CsCu$_2$X$_3$ (X= Cl, Br, I) \cite{Roccanova2019,Lin2019,Li2020}. It has been reported that these classes of halides exhibit high photoluminescence quantum efficiencies (PLQEs),
as well as wide, tunable ranges of emission wavelengths. These highly desirable properties can be attributed to the presence of self-trapped
excitons \cite{Luo2018,Li2019,Lian2020}: thanks to the reduced dimensionality and the
relatively soft crystal lattices, the excitons in these materials are strongly self-trapped, which results in a large Stokes shift and a broad spectrum of
photoemission \cite{ref10,ref11}. Clearly, a theoretical understanding of the excitons in these materials is important for improving the PLQE.
A complete description including the coupling of electronic and lattice degrees of freedom would be a highly challenging task;
however, even without lattice relaxation effects, an accurate first-principles study of excitons in Cu halides will be a crucial
step towards a more complete picture of the optical properties in these materials.

Due to the weakened Coulomb screening, excitons in low-dimensional Cu halides are much more strongly bound than in bulk \cite{Rb2CuBr3_ref2,Roccanova2019,Lin2019}. The excitonic binding energy is reported to be several hundreds of meV \cite{Roccanova2019,Rb2CuBr3_ref2}, which is one to two orders of magnitude larger than those in the popular 3D lead halide based perovskites (CsPbX$_3$ \cite{Miyata2015,nl6b01168,C6RA17008K} and CH$_3$NH$_3$PbX$_3$ \cite{Bokdam2016,JoshuaAndre2019}).
Excitons are collective excitations within a many-body system, and thus cannot be properly described within a single-quasiparticle (QP) picture.
To study the excitons, the standard method is via many-body perturbation theory, using the $GW$ approximation for the QP states \cite{Reining2017}
and the Bethe-Salpeter equation (BSE)
for the optical spectra \cite{Rohlfing2000,Reining2016,yuchen2016}. However, this approach becomes extremely expensive for large supercells, which makes it unaffordable for defective or lattice-distorted systems.

An alternative approach is density functional theory (DFT) and its dynamical counterpart, time-dependent density functional theory (TDDFT) \cite{TDDFT84,Onida2002,Reining02,Ullrich2012}. A very widely used generalization of (TD)DFT \cite{Seidl1996,Gorling1997,Baer2018,Garrick_Kronik2020} is based on hybrid functionals, where a fraction of nonlocal Fock exchange is combined with (semi)local exchange and correlation \cite{Becke1993,Becke1996,Perdew1996b,Stephens1994}.
The prefactor of the Fock exchange part can be determined either by empirical fitting \cite{Adamo1999,Heyd2003,Krukau2006}, or by using a nonempirical value dependent on the dielectric function \cite{Skone2014,Skone2016,Chen2018}.
In ground-state DFT, hybrid xc functionals have gained increased popularity for calculating electronic band structures, since they offer a practical solution to DFT’s band-gap problem \cite{Cora2004,Heyd2005,Gerber2007,Henderson2011,Jain2011,Marques2011,Matsushita2011,Moussa2012,Friedrich2012,Gerosa2015,Hinuma2017,HongJiang2020}.
Although hybrid functionals have been applied for ground-state band structures in perovskites \cite{Franchini_2014,Jiang2018,Bischoff2019,Zu2019,Du2020}, there have so far been only
few applications of hybrid functionals for the optical spectra of perovskites \cite{Jiuyu2019}.

\begin{figure}
	\centering
\includegraphics[width=8.0 cm]{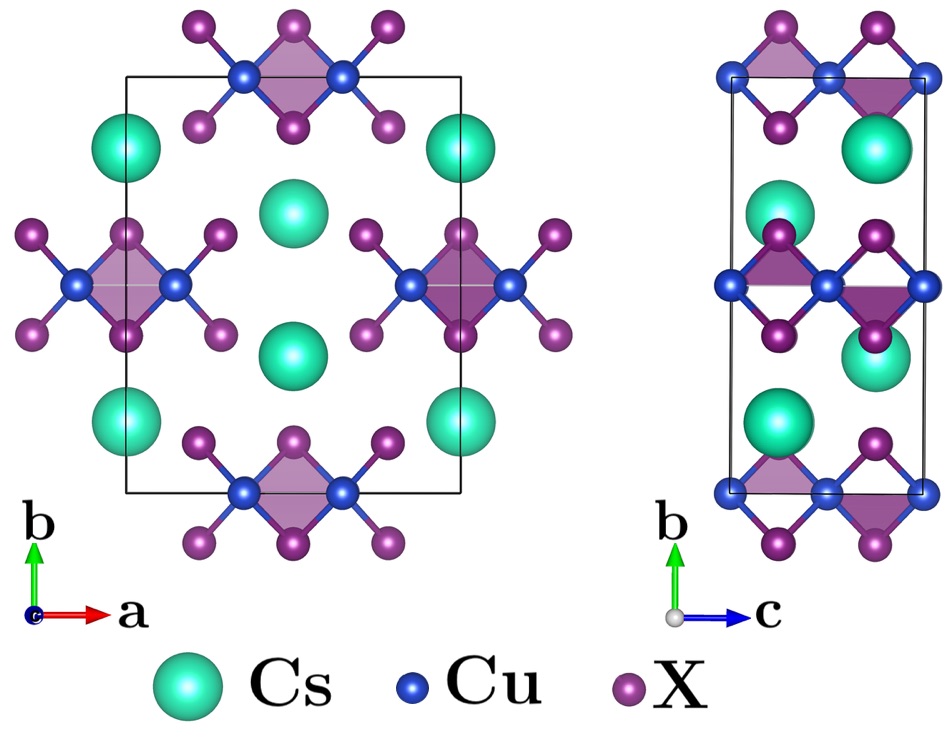}
     \caption{Crystalline geometry of CsCu$_2$X$_3$ (X= Cl, Br, I). The lattice features 1D chains of $\rm [Cu_2X_3]^-$ units along the $c$ axis, separated by rows of $\rm Cs^+$ ions \cite{Roccanova2019}. One orthorhombic unit cell is shown, containing 4 Cs, 8 Cu and 12 halogen atoms.}
     \label{fig:str}
\end{figure}

In this paper, we explore the performance of hybrid functionals for calculating the optical properties, in particular the excitons, of
CsCu$_2$X$_3$ (X= Cl, Br, I) (see Fig.~\ref{fig:str}).  As mentioned, this class of low-dimensional perovskites has excitonic binding energies over 100 meV, which presents
an ideal test case to examine the performance of hybrid TDDFT for excitonic effects.
Detailed experimental results for the optical absorption spectra and exciton binding energies of CsCu$_2$X$_3$ are not available at present;
therefore, we base our assessment of hybrid TDDFT on a comparison with reference results from state-of-the-art $GW$-BSE; to our knowledge,
QP band structures and optical properties from $GW$-BSE have not been previously reported for these materials.

We apply both empirical and nonempirical hybrid functionals,
in which the admixture of nonlocal exchange is fixed or material-dependent, respectively.
Comparison between different hybrid functionals and the BSE
shows that nonempirical hybrid functionals that depend on the dielectric properties of the material yield accurate optical spectra for CsCu$_2$X$_3$.
This suggests that hybrid functionals are very well suited for calculating the optical properties in all-inorganic perovskites,
and more generally for calculations with large unit cells and high-throughput computations.

This paper is organized as follows: in Sec. \ref{sec:II} we give an overview of the theoretical background, with an emphasis on the different
TDDFT hybrid functionals. We also discuss computational details. The results for the optical spectra and excitonic properties of CsCu$_2$X$_3$
are presented in Sec. \ref{sec:III}. Section \ref{sec:IV} then gives our conclusions and an outlook.

\section{Theory and Computational Details}\label{sec:II}

\subsection{A Brief Overview of Hybrid Functionals}

In ground-state DFT, the central idea of hybrid functionals is to write the exchange-correlation (xc)
energy functional as \cite{Becke1993,Becke1996,Perdew1996b,Stephens1994,Adamo1999}
\begin{equation}\label{Exc_hybrid}
E_{\rm xc}^{\rm hybrid} = a E_{\rm x}^{\rm exact} + (1-a)E_{\rm x}^{\rm sl} + E_{\rm c}^{\rm sl} \:,
\end{equation}
where $E_{\rm x}^{\rm exact}$ is the exact nonlocal (Fock) exchange energy functional, and $E_{\rm x,c}^{\rm sl}$ are approximate (semi)local  (sl) exchange and correlation functionals,
respectively. $E_{\rm xc}^{\rm hybrid}$ contains a parameter $a$ which mixes exact exchange with (semi)local exchange.
There are hybrid functionals with two or more mixing parameters \cite{Stephens1994}, but for simplicity we will not include them in our discussion.
In practice, hybrid functionals are normally used within the so-called generalized Kohn-Sham approach \cite{Garrick_Kronik2020}.

One may distinguish different types of hybrid functionals. In {\em global} hybrids, the fraction $a$ of exact exchange is a constant throughout the system.
In {\em local} hybrids, the parameter $a$ depends on position \cite{Maier2018}.
The {\em range-separated} hybrid functionals \cite{Baer2010}, by contrast, consider the xc interactions of long range and short range separately. The most popular
way of doing this is by partitioning the Coulomb interaction as
\begin{equation}\label{range_sep}
\frac{1}{r} = \frac{\alpha+\beta {\rm erf}(\gamma r)}{r} + \frac{1-[\alpha+\beta {\rm erf}(\gamma r)]}{r} \:,
\end{equation}
where the first (long-range) term enters the Fock exchange, and the second (short-range) term enters the (semi)local exchange \cite{Iikura2001},
or vice versa \cite{Janesko2009}. Typical examples for the global and range-separated hybrid functionals are PBE0 \cite{Adamo1999} (with simply $a=0.25$) and HSE06
\cite{Heyd2003}, respectively. Although the parameters in both functionals are determined empirically, PBE0 and HSE06, as well as other empirical hybrid functionals, have been widely used for periodic solids, mainly for the resulting improvement of band gaps
\cite{Heyd2005,Henderson2011,Jain2011,Marques2011,Matsushita2011,Moussa2012,Skelton2020}. The range separation parameters can also be determined (or ``tuned'')
using certain exact conditions, which has been quite successful for molecules and molecular crystals \cite{Baer2010,Refaely2015}.

On the other hand, the ongoing effort to construct nonempirical hybrid functionals for solids has made remarkable progress.
A class of dielectric-dependent hybrid functionals  (DDHs) have been proposed \cite{Marques2011,Gerosa2015,Skone2014,Skone2016,Chen2018}, where the dielectric function is used to determine the parameters in Eqs. (\ref{Exc_hybrid}) or (\ref{range_sep}). Since the dielectric function can be calculated from the random phase approximation (RPA), the resulting hybrid functionals are made nonempirical. For solids with a finite gap, the screening of the long-range tail of the Coulomb interaction is proportional to the inverse of the static dielectric constant ($\varepsilon_\infty^{-1}$).
It is intuitive to identify the parameter $a$ in Eq. (\ref{Exc_hybrid}) with $\varepsilon_\infty^{-1}$, which directly leads to a global DDH. For range-separated DDHs \cite{Skone2016,Chen2018}, using the same long-range limit of Coulomb interaction, one obtains $\alpha+\beta=\varepsilon_\infty^{-1}$. The only remaining parameter,
the range-separation parameter $\gamma$, is then fitted to reproduce either the fundamental band gap or the finite-range dielectric function of the system.
In addition, the parameters in both global and range-separated DDHs can be optimized via an iterated procedure to obtain a self-consistent dielectric constant (or dielectric function) \cite{Skone2014,Bischoff2019}.
In the following, we refer to DDHs where the dielectric function is calculated using the RPA with (nonhybrid) DFT wave functions
as {\it single-shot} hybrid functionals, analogous to the widely used single-shot $G_0W_0$ approach.

Applications of hybrid TDDFT for optical spectra in semiconductors and insulators have begun to emerge
\cite{Paier2008,Bernasconi2011,Tomic2014,Webster2015,Refaely2015,Kronik2016,Wing2019,Ramasubramaniam2019,Lewis_Sahar2020,Jiuyu2019,Yang2015,Alexey2020}.
For instance, Lewis {\em et al.} recently showed that range-separated hybrids with an empirical parameter agree well with $GW$+BSE results for gallium nitride
with defects \cite{Lewis_Sahar2020}. Furthermore, it was recently reported that DDHs yield band gaps of good accuracy for inorganic metal-halide perovskites \cite{Bischoff2019};
this suggests that DDHs may be suitable for the optical properties of perovskites. We will see below that this is indeed the case.

\subsection{Connection between BSE and hybrid functionals for excitons in solids}

From the point of view of generalized (TD)DFT, hybrid functionals bridge the density-based (TD)DFT and the orbital-based many-body theory via a fraction of Fock exchange,
while representing the $N$-electron wave function as a single Slater determinant  with single-particle orbitals $\varphi_i$ \cite{Garrick_Kronik2020,Baer2018}.
For ground states, global DDHs connect to the static approximation of $GW$, known as the static COulomb Hole plus Screened EXchange (COHSEX) \cite{Skone2014}. By adopting $a=\varepsilon_\infty^{-1}$, the long-ranged Fock exchange in Eq. (\ref{Exc_hybrid}) corresponds to the statically screened exchange (SEX) self-energy:
\begin{equation}\label{SEX}
\begin{split}
\Sigma_{\rm SEX}(\bfr,\bfr') & = -\sum_{i=1}^{N}\varphi_i(\bfr)\varphi^*_i(\bfr')W(\bfr,\bfr')   \\
& \approx -\varepsilon_\infty^{-1}\sum_{i=1}^{N}\frac{\varphi_i(\bfr)\varphi^*_i(\bfr')}{|\bfr-\bfr'|} \:.
\end{split}
\end{equation}
Here, the screened Coulomb potential $W$ is given by
\begin{equation}
W(\bfr,\bfr') = \int d\bfr'' \: \frac{\varepsilon^{-1}(\bfr,\bfr')}{|\bfr' - \bfr'|} \:.
\end{equation}
For the range-separated DDHs, the error function $\mbox{erf}(\gamma|\bfr-\bfr'|)$
offers a good model for the dielectric function $\varepsilon^{-1}(\bfr,\bfr')$ in 3D solids.
In reciprocal space, $\varepsilon^{-1}$ thus becomes \cite{Skone2016,Chen2018}
\begin{equation}\label{model_eps}
\varepsilon^{-1}(q) = \varepsilon_\infty^{-1}+(1-\varepsilon_\infty^{-1}) (1- e^{-q^2/4\gamma^2})\:.
\end{equation}
Thus, $W(\bfr,\bfr')$ can be evaluated using model dielectric functions in Eq. (\ref{model_eps}),
thereby connecting range-separated DDHs to the COHSEX via Eq. (\ref{SEX}) \cite{Skone2016,Chen2018}.

For excited states, one finds that generalized TDDFT is connected to the BSE in a similar way. Taking the Tamm-Dancoff approximation with momentum transfer $\mathbf{q}=0$ \cite{Sander15,Byun2017,YoungMoo2017}, the BSE can be simply expressed as
\begin{equation}\label{TDA}
\left[ (E_{c,\mathbf{k}} - E_{v,\mathbf{k}'})\delta_{vv'}\delta_{cc'}\delta_{\mathbf{k}\mathbf{k}'} + K^{\mathrm{BSE}}_{cv\mathbf{k},c'v'\mathbf{k}'}   \right] \mathbf{Y}_n =  \omega_n \mathbf{Y}_n.
\end{equation}
Here, $v$ denotes occupied valence bands, $c$ denotes unoccupied conduction bands, the $E_\mathbf{k}$ are QP energies, $\mathbf{Y}_n$ are the $n$th eigenvectors, and $\omega_n$ is the $n$th excitation energy. Equation (\ref{TDA}) features
the coupling matrix (or BSE kernel) $K^{\mathrm{BSE}} = K^{\mathrm{d}} + K^{\mathrm{x}}$. The first part, $K^{\rm d}$,
is the direct interaction, corresponding to the Hartree part in TDDFT. The second part is the screened exchange kernel,
\begin{equation}
\label{eqn:Kd}
\begin{aligned}
K^{\mathrm{x}}(\mathbf{q},\omega) & =  \frac{2}{V_{\rm cell}}\sum_{\mathbf{G},\mathbf{G'}} W_{\mathbf{G},\mathbf{G'}}(\mathbf{q},\omega) \delta_{\mathbf{q},\mathbf{k}-\mathbf{k}'}\\
& \times \bra{c\mathbf{k}} e^{i(\mathbf{q}+\mathbf{G})\mathbf{r}}\ket{c'\mathbf{k}'} \bra{v'\mathbf{k'}} e^{-i(\mathbf{q}+\mathbf{G'})\mathbf{r}}\ket{v\mathbf{k}} .
\end{aligned}
\end{equation}
Here, $W_{\mathbf{G},\mathbf{G'}}(\mathbf{q},\omega)$ is the dynamically screened Coulomb interaction in reciprocal space,
\begin{equation}
\label{eqn:W}
 W_{\mathbf{G},\mathbf{G'}}(\mathbf{q},\omega)  =
        -\frac{4\pi \varepsilon^{-1}_{\mathbf{G},\mathbf{G'}}(\mathbf{q},\omega)}{|\mathbf{q}+\mathbf{G}||\mathbf{q}+\mathbf{G'}|},
\end{equation}
where in standard BSE $\varepsilon^{-1}_{\mathbf{G},\mathbf{G'}}(\mathbf{q},\omega)$ is obtained via RPA, as mentioned before. For most practical BSE calculations,
the frequency dependence of the dielectric function is ignored, i.e., one
uses $\varepsilon^{-1}_{\mathbf{G},\mathbf{G'}}(\mathbf{q},\omega=0)$ \cite{Rohlfing98,Benedict98,bseGaN}.

Similar to the approximations for $GW$, one can replace the RPA dielectric function by a model dielectric function $\varepsilon^{-1}_{\rm m}(q)$
or just a simple parameter, both of them diagonal
in the reciprocal lattice vectors $\bfG,\bfG'$:
\begin{equation}
\label{m-BSEandSXX}
 \varepsilon^{-1}_{\mathbf{G},\mathbf{G'}}(\mathbf{q},0)  \longrightarrow
 \left\{\begin{aligned}
       \varepsilon^{-1}_{\rm m}(q) \delta_{\mathbf{G},\mathbf{G'}} & \qquad 	\textrm{(m-BSE)}\\
        a \delta_{\mathbf{G},\mathbf{G'}}	 &\qquad \mathrm{(SXX).}
       \end{aligned}
\right.
\end{equation}
Here, m-BSE stands for BSE with a model dielectric function, and SXX stands for screened exact exchange \cite{Yang2015,Jiuyu2019}.
If $a=1$, SXX reduces to time-dependent Hartree-Fock.

Combining SXX and (semi)local exchange and correlation thus leads to a hybrid kernel for generalized TDDFT,
analogous to Eq. (\ref{Exc_hybrid}):
\begin{equation}\label{K_hybrid1}
K_{\rm xc}^{\rm hybrid1} = K^{\rm SXX} + (1- a) K_{\rm x}^{\rm sl} +  K_{\rm c}^{\rm sl}\:,
\end{equation}
where $K_{\rm x}^{\rm sl}$ and $K_{\rm c}^{\rm sl}$ are the adiabatic (semi)local exchange and correlation kernels, respectively.
Replacing the BSE kernel $K^{\rm BSE}$ with the hybrid kernel in Eq. (\ref{TDA}) defines our generalized TDDFT scheme for optical excitations in solids.
In the following, we choose the adiabatic local-density approximation (ALDA) kernels to construct  the hybrid kernel, i.e., we set
\begin{eqnarray}\label{K_ALDA}
K_{\rm x,c}^{\rm sl}(\mathbf{q})  &=&  \frac{2}{V_{\rm cell}}\lim_{{\bf q}\to 0}\sum_{\mathbf{G},\mathbf{G'}} f_{{\rm x,c},{\bf G} {\bf G'}}^{\rm ALDA}({\bf q})
\\
&\times&
\bra{c\mathbf{k}} e^{i(\mathbf{q}+\mathbf{G})\mathbf{r}}\ket{v\mathbf{k}} \bra{v'\mathbf{k'}} e^{-i(\mathbf{q}+\mathbf{G}')\mathbf{r}}\ket{c'\mathbf{k}'} ,
\nonumber
\end{eqnarray}
where $f_{{\rm x,c},{\bf G} {\bf G'}}^{\rm ALDA}({\bf q})$ are the local frequency-independent exchange and correlation kernels in ALDA \cite{Ullrich2012,Byun_2020}.

The expressions Eq. (\ref{Exc_hybrid}) and Eq. (\ref{K_hybrid1}) are constructed following the standard TDDFT hybrid functional form.
However, in extended many-body systems, not only the exchange but also the correlation is subject to dielectric screening; this can, for instance,
be explicitly seen in the COH part of the COHSEX approximation for the QP self-energy \cite{Bruneval2006}.
Hence, the (semi)local correlation energy, $E^{\rm sl}_{\rm c}$, should also be screened in
order to avoid overcounting of correlation. This leads us to define a second form of hybrid functional,
\begin{equation}\label{K_hybrid2}
K_{\rm xc}^{\rm hybrid2} = K^{\rm SXX} + (1-a) K_{\rm xc}^{\rm sl} \:,
\end{equation}
which was already proposed earlier, and was found to compare well with the BSE \cite{Jiuyu2019}.

In the following, we will assess three hybrid kernels based on the definitions in Eq. (\ref{K_hybrid1}) and Eq. (\ref{K_hybrid2}).
The first one, LDA0, is built as a traditional empirical hybrid functional in analogy with PBE0 \cite{Adamo1999}, by setting $a=0.25$ in Eq. (\ref{K_hybrid1}).
The other two, DDH1 and DDH2, are built by identifying $a$ with the RPA inverse dielectric constant $\varepsilon_{0,0}^{-1}$ in Eqs. (\ref{K_hybrid1}) and (\ref{K_hybrid2}),
respectively.
A detailed comparison of the hybrid kernels will be given in Section \ref{sec:IIIB}.

\subsection{Computational Details}

For CsCu$_2$X$_3$,  calculating the crystal structure with relaxed atomic positions using pure DFT, hybrid functionals or even $GW$ can lead to errors compared
to the experimental lattice constants and geometries \cite{Du2020}, which affects the calculated band gaps in uncontrolled ways.
To eliminate this problem, we simply adopted the experimental crystal structures,
lattice constants and atomic positions for all three materials CsCu$_2$X$_3$ \cite{Roccanova2019}.

To calculate Kohn-Sham band structures, we used the LDA and PBE \cite{Perdew1996} functionals in the Quantum Espresso package \cite{QE-2017}, based on the plane-wave basis with cutoff energy of 85 Ry. We adopted optimized norm-conserving Vanderbilt pseudopotentials \cite{ONCVPSP1,ONCVPSP2} for all the elements with both LDA and PBE, with the exception
of using the Troullier-Martins pseudopotential for Cs with LDA \cite{Troullier1991}. The validation of this mixed pseudopotential setting is provided by comparing the band structures from LDA and PBE functionals (see Supplemental Material \cite{supp}).

Starting from the Kohn-Sham band structures, we then used the Yambo code \cite{Yambo2019} to obtain dielectric functions and QP band structures. The RPA dielectric functions were calculated with 300 bands and 600 $\bf G$-vectors. For the QP band structures, we used the single-shot $G_0W_0$ scheme with at least 400 bands. Although the dielectric function, as well as $GW$+BSE, are introduced within the static approximation, the $\omega$-dependent dynamic effects of the dielectric screening may have a nonnegligible influence on the electron dynamics \cite{Hybertsen1986,Godby1989,Marini2003,GPP1}.
In this work, the generalized-plasmon-pole model \cite{Godby1989} was applied to account for the dynamical effects. Thanks to the flexible definition of $K^{\rm SXX}$, it is compatible with dynamically corrected RPA inverse dielectric constants ($\varepsilon^{-1}_{0,0}$). In the following, such definition of $\varepsilon^{-1}_{0,0}$ is used, including building the DDHs, unless explicitly stated otherwise.

A double-grid was adopted, which includes a $4\times3\times8$ $\Gamma$-centered $\bfk$-point mesh plus 3000 random interpolated $\bf k$-points. This enabled us to apply the random integration method, as well as an inversion solver for the BSE-type equations \cite{Yambo2019,Kammerlander2012}.  To build the BSE-type kernels,
we used 56 valence bands and 20 conduction bands for CsCu$_2$Cl$_3$. The corresponding numbers of valence and conduction bands are 44 and 10 for CsCu$_2$Br$_3$ and 18 and 14
for CsCu$_2$I$_3$. A broadening of 0.1 eV was used to calculate the optical spectra.

\begin{figure}
	\centering
\includegraphics[width=8.0 cm]{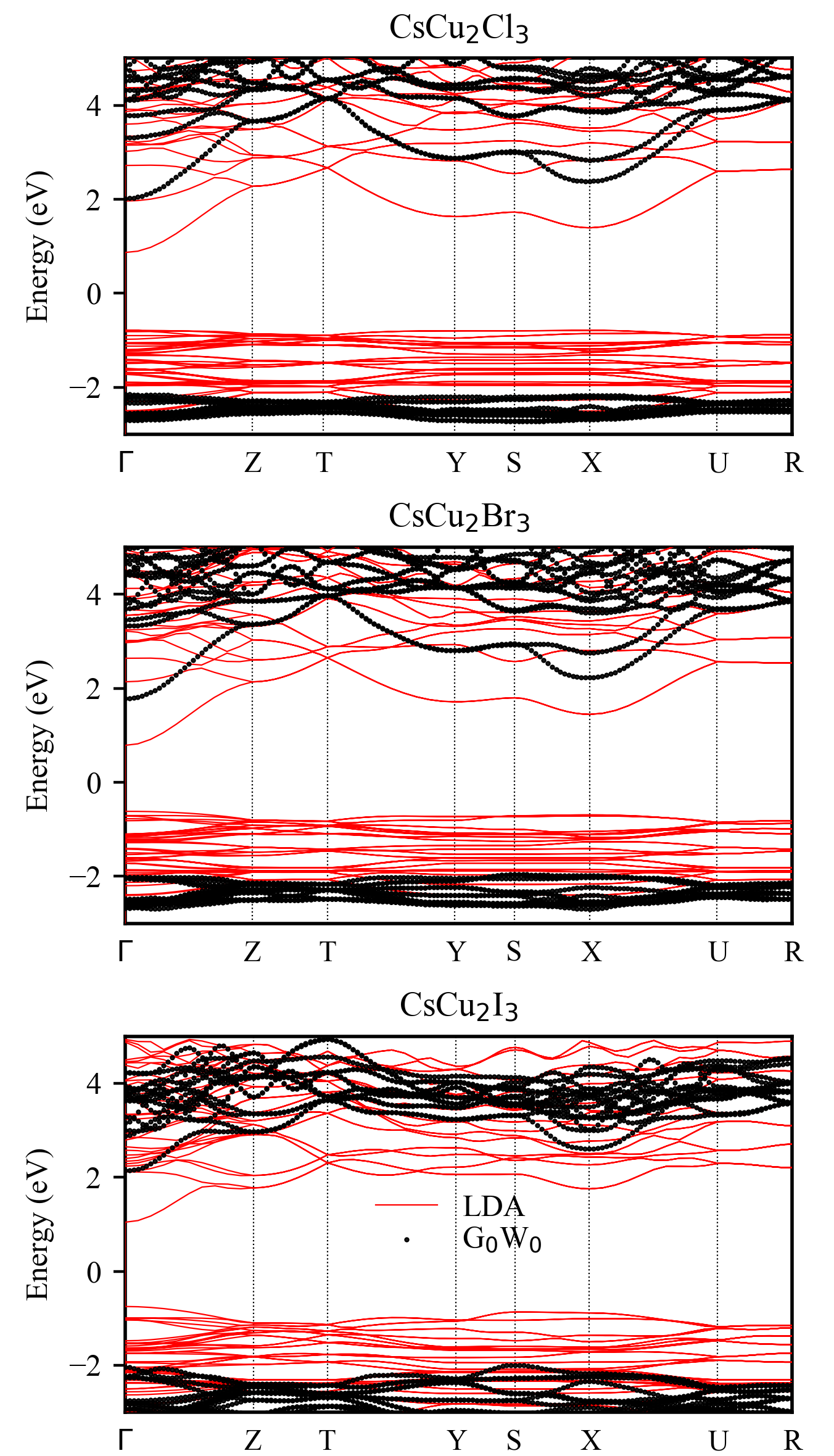}
     \caption{Band structures of CsCu$_2$X$_3$, calculated with LDA and $G_0W_0$@LDA.}
     \label{fig:bands}
\end{figure}

We performed convergence tests regarding the $\bfk$-point grid and number of bands included in the $G_0W_0$ calculations, and found that
the choices detailed above lead to well-converged results for band gaps, exciton binding energies and optical spectra (see Supplemental Material \cite{supp}).

\section{Results and Discussion} \label{sec:III}

\subsection{Band Structures and band gaps}

We begin by comparing the band structures of the Cs-Cu halides calculated with different approaches, namely,  Kohn-Sham band structures from LDA and PBE,
and single-shot $GW$ band structures on top of LDA and PBE ($G_0W_0$@LDA and $G_0W_0$@PBE).

Figure \ref{fig:bands} shows the band structures of CsCu$_2$X$_3$ obtained by LDA and $G_0W_0$@LDA. As expected, there are important differences,
most notably a pronounced opening up of the band gaps caused by the QP corrections.
Along with the enlarged band gaps, the conduction bands in all three materials are significantly altered in $G_0W_0$, especially around the X point. In addition, the valence bands are strongly affected as well: for the case of CsCu$_2$I$_3$, this is most apparent for the highest occupied
valence bands between the $\Gamma$ and Z points, as well as for the valence bands around the S point. Overall, these differences in the band structures
suggest that a simple scissors operator approach (which would not change the shape of the LDA bands, just shift them rigidly) is not sufficiently accurate for calculating
the optical properties in these materials.

In the Supplemental Material \cite{supp}, we show that the band structures obtained by LDA and $G_0W_0$@LDA are very similar to the PBE and $G_0W_0$@PBE band structures,
respectively, except for some differences of the band gaps, see below. In view of this, will use the $G_0W_0$@LDA band structures as input for the BSE and TDDFT calculations
of the optical spectra in the following sections.

   \begin{table}
     \caption{Band gaps of CsCu$_2$X$_3$, calculated with DFT and $G_0W_0$. All energies are in eV. }
     \label{tbl:gap}
     \begin{tabular*}{0.48\textwidth}{@{\extracolsep{\fill}}lccccc} 
     \hline
    					     & LDA  	& $G_0W_0$@LDA& PBE &$G_0W_0$@PBE & PBE0\footnotemark[1] \\\hline
   CsCu$_2$Cl$_3$   & 1.65  		&4.20  	  		&  1.78			 & 3.92  	& 4.29\\
   CsCu$_2$Br$_3$   & 1.40		&3.63		 		&  1.59 			 & 3.47   	& 3.88\\
   CsCu$_2$I$_3$     &  1.79		&4.03  			&  1.98  			 & 3.83   	&  3.93 \\\hline
     \end{tabular*}
\footnotetext[1]{from Ref. \cite{Du2020}}
   \end{table}

The calculated band gaps are listed in Table~\ref{tbl:gap}, together with PBE0 results from Ref. \cite{Du2020}.
As already pointed out, both LDA and PBE gaps are significantly smaller than the corresponding $G_0W_0$ gaps. Unfortunately, no experimental results for the QP gaps
$E_{\rm g}^{\rm QP}$ in CsCu$_2$X$_3$ are available. In Section \ref{sec:IIIB} we will discuss the optical gaps $E_{\rm g}^{\rm opt}$ in some detail; due to excitonic
effects, the optical gap is a lower limit to the QP gap. As seen in Table \ref{tbl:optic}, the experimental optical gaps are on the order of 3.8 eV \cite{Roccanova2019,Lin2019}, which clearly shows that LDA and PBE dramatically underestimate the QP gaps.

After QP correction, the $G_0W_0$ gaps are drastically enhanced over the corresponding Kohn-Sham gaps, by about 2 eV. While LDA gives smaller band gaps than PBE, the
$G_0W_0$@LDA gap is larger than the $G_0W_0$@PBE gap for all materials under study. We note that the band gaps obtained with PBE0  \cite{Du2020}
are comparable to $G_0W_0$, which indicates the capability of hybrid functionals for predicting band gaps in perovskites.

We also implemented the DDH1 and DDH2 functionals to calculate the band gaps; to reduce computational cost, the calculations
were performed single-shot rather than self-consistently. It is found that single-shot DDH1 and DDH2 tend to somewhat overestimate
the band gap compared to $G_0W_0$. These results are discussed in more detail in the Supplemental Material \cite{supp}.

A special case here is CsCu$_2$Br$_3$, where the $G_0W_0$ gaps are smaller than the experimental optical gap \cite{Roccanova2019}.
Possible explanations for this underestimation could be missing core electron effects \cite{Bischoff2019} or
$G_0W_0$ starting point dependence \cite{Filip2014,Leppert2019} (see Supplemental Material \cite{supp} for further discussion).

On the other hand, the PBE0 gap is better compatible with the experimental $E_{\rm g}^{\rm opt}$; however, the PBE0 calculation was done with the relaxed 
lattice structure rather than the experimental geometry \cite{Du2020}, and the band gap is sensitive to changes of the atomic positions.

\subsection{Optical spectra via the BSE} \label{sec:IIIB}

\begin{figure}
	\centering
\includegraphics[width=8.0 cm]{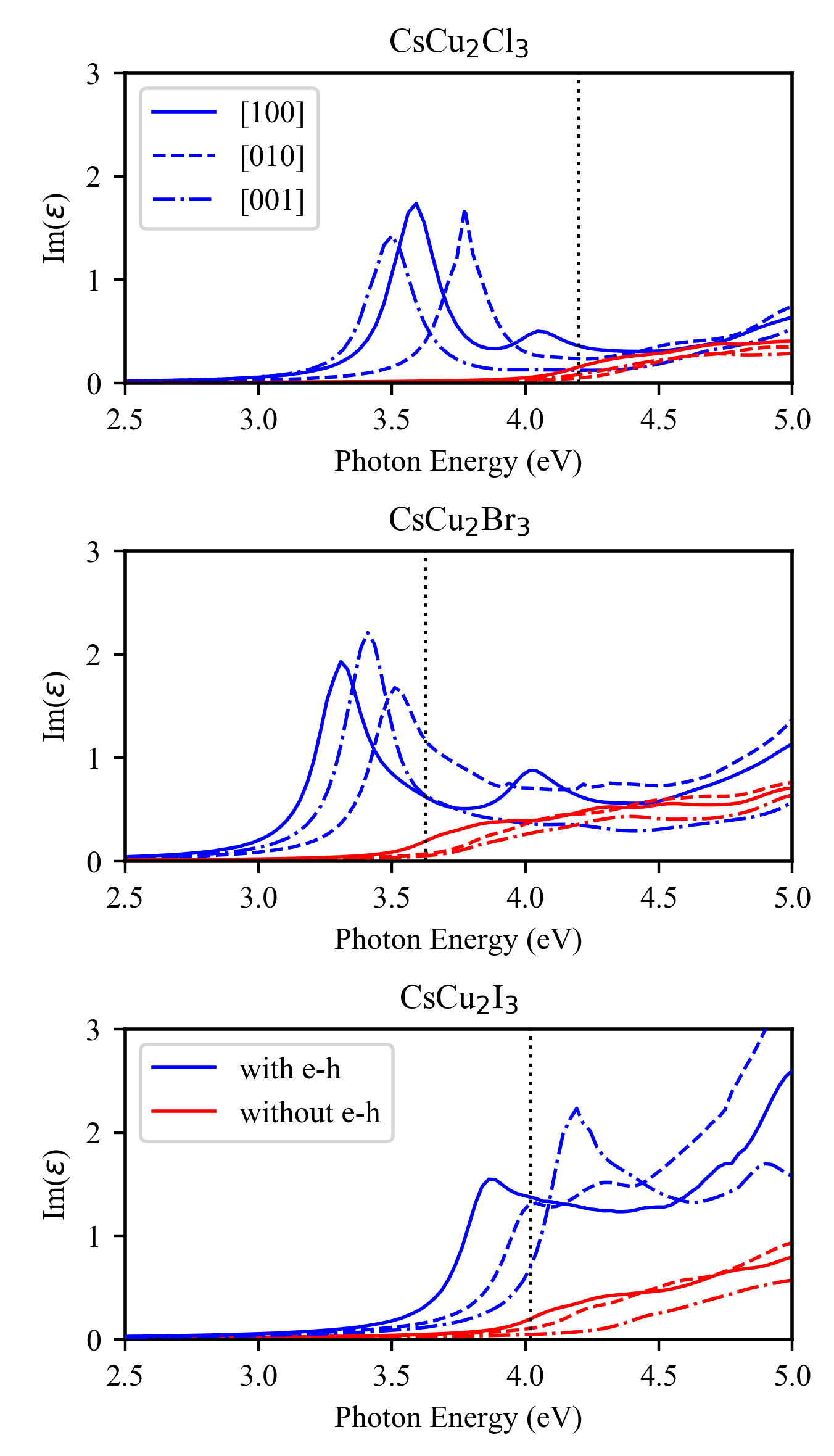}
     \caption{Optical spectra Im($\varepsilon$) calculated with $G_0W_0$+RPA (red lines) and $G_0W_0$+BSE (blue lines). The vertical dotted lines indicate
     the positions of the $G_0W_0$@LDA band gaps.}
     \label{fig:bse_spc}
\end{figure}

We now discuss the optical properties of the Cs-Cu halides, beginning with the $G_0W_0$+BSE results.
We compare the spectra associated with light polarization along the [100], [010], and [001] directions, which highlights the anisotropic,
quasi-1D structure of CsCu$_2$X$_3$, see Fig. \ref{fig:str}. The excitonic biding energies are obtained via the standard definition, i.e.
$E_{\rm b} = E^{\rm QP}_{\rm g} - E_{\rm exc}$, where $E_{\rm exc}$ is the excitation energy of the lowest bright exciton along corresponding direction.
The calculated excitonic binding energies are given in Table~\ref{tbl:optic},
and the $G_0W_0$+RPA  and $G_0W_0$+BSE optical spectra (imaginary part of the dielectric function) are displayed in Fig.~\ref{fig:bse_spc}.

   \begin{table}
     \caption{Excitonic binding energies ($E_{\rm b}$, in meV) calculated using BSE, RPA inverse dielectric constants ($\varepsilon^{-1}_{0,0}$) along different directions, as well as calculated and experimental (in parentheses) optical gaps ($E^{\rm opt}_{\rm g}$, in eV).}
     \label{tbl:optic}
     \begin{tabular*}{0.48\textwidth}{@{\extracolsep{\fill}}lccccccc} 
     \hline
					   & 	\multicolumn{2}{c}{[100]} &	\multicolumn{2}{c}{[010]} 	& \multicolumn{2}{c}{[001]} &   \\
					     & $E_{\rm b}$ & $\varepsilon^{-1}_{0,0}$ & $E_{\rm b}$ & $\varepsilon^{-1}_{0,0}$ & $E_{\rm b}$ & $\varepsilon^{-1}_{0,0}$ & $E^{\rm opt}_{\rm g}$\\\hline
   CsCu$_2$Cl$_3$   &  611 	&0.35   &  425	&  0.37 & 701 &  0.31& 3.50 (3.89\footnotemark[1]) \\
   CsCu$_2$Br$_3$   & 321	&0.30	 &  120 	&	0.31 & 219 & 0.28 & 3.32 (3.89\footnotemark[1])\\
   CsCu$_2$I$_3$     &  156	&0.26   &  51  &	0.26	 &   32	 &  0.25 & 3.86 (3.71\footnotemark[1], 3.78\footnotemark[2]) \\\hline
     \end{tabular*}
     \footnotetext[1]{from Ref. \cite{Roccanova2019}}
     \footnotetext[2]{from Ref. \cite{Lin2019}}
   \end{table}

As shown in Table~\ref{tbl:optic},  the $E_{\rm b}$ in CsCu$_2$Cl$_3$ along [100], [010], and [001] are 611 meV, 425 meV, and 701 meV, respectively, which is almost ten times larger than in the widely studied lead halide perovskites,  CsPbX$_3$ and CH$_3$NH$_3$PbX$_3$ \cite{Miyata2015,nl6b01168,C6RA17008K,Bokdam2016,JoshuaAndre2019}.
These large binding energies are associated with distinct excitonic optical absorption peaks that are separated from the continuum part of the spectrum,
as observed in the upper panel of Fig.~\ref{fig:bse_spc}. In the BSE optical spectra,
strong absorption peaks appear below the QP band gap for all three directions; these peaks are completely missing in RPA.

Comparing the $E_{\rm b}$ of the three materials in Table~\ref{tbl:optic}, one finds that the excitons become more and more weakly bound going from $\rm X=Cl$ to Br and then
to I. From Fig. \ref{fig:bse_spc} we see that the excitonic peaks move closer and closer to the QP band gaps for CsCu$_2$Br$_3$ and CsCu$_2$I$_3$. In particular, there is no
longer a separate excitonic peak along [010] and [001] for CsCu$_2$I$_3$. The corresponding $E_{\rm b}$ are 51 meV and 32 meV, which is of the same magnitude as
$E_{\rm b}$ in CsPbX$_3$ and CH$_3$NH$_3$PbX$_3$, and similar to those in many binary semiconductors such as GaN or ZnO. In all of those cases, one observes that
the excitonic peaks merge with the onset of the absorption continuum above the QP gap.

The main reason for the decreasing $E_{\rm b}$ from CsCu$_2$Cl$_3$ to CsCu$_2$I$_3$ is the increasing dielectric screening. As shown in Table~\ref{tbl:optic}, a decreasing
trend of RPA $\varepsilon^{-1}_{0,0}$ is found from CsCu$_2$Cl$_3$ to CsCu$_2$I$_3$ for each direction. This reflects an enhancement of the overall dielectric screening as the X ions evolve from Cl to I, which is also found in CsBX$_3$ (B = Ge, Sn, Pb and X = Cl, Br, I) \cite{Bischoff2019}.
Moreover, the experimentally measured binding energies of trapped excitons show the same trend \cite{Roccanova2019}, though they differ from the $E_{\rm b}$ of free
excitons that are considered here.

There is a strong anisotropy of the excitons in CsCu$_2$X$_3$: $E_{\rm b}$ along [100] is significantly larger than that along [010] for each CsCu$_2$X$_3$.
On the other hand, the relation of the [001] exciton to the [100] and [010] excitons changes with X: $E_{\rm b}$ along [001] is the largest of the three in CsCu$_2$Cl$_3$, but then becomes the intermediate and the smallest one in CsCu$_2$Br$_3$ and CsCu$_2$I$_3$, respectively. This change can also be nicely seen for the [001]
absorption peak in the optical spectra in Fig. \ref{fig:bse_spc}. These anisotropy effects cannot be explained by the macroscopic dielectric screening;
for example, in CsCu$_2$Cl$_3$, $E_{\rm b}$ along [100] is smaller than along [001], but $\varepsilon^{-1}_{0,0}$ along [100] is larger than along [001].
A better explanation would be the strong local-field effects in these ionic crystals.

Besides the absence of distinct excitonic peaks below the band gap, there are additional differences between the BSE and RPA spectra. First, the continuum absorption
above the band gap is enhanced by including electron-hole interactions, especially for CsCu$_2$I$_3$. Secondly, the RPA spectrum in each material always has the highest
absorption edge along the [001] direction. In other words, the change of the absorption edges along [001] discussed in last paragraph is not observed in RPA.
Clearly, the RPA is inadequate for calculating the optical properties of CsCu$_2$X$_3$.

Lastly, in the rightmost column of Table~\ref{tbl:optic} we compare the calculated optical gap with experiment \cite{Roccanova2019,Lin2019}.
The optical gap is defined as $E^{\rm opt}_{\rm g} = E^{\rm QP}_{\rm g}-E_{\rm b} $, where
$E^{\rm QP}_{\rm g}$ is here the fundamental $G_0W_0$@LDA band gap and $E_{\rm b}$ is the binding energy of the lowest energy exciton among all three directions.
The calculated $E^{\rm opt}_{\rm g}$ of CsCu$_2$I$_3$ is in good agreement with experiment, but it is somewhat too low for CsCu$_2$Cl$_3$ and CsCu$_2$Br$_3$ (by 0.39 and 0.57 eV, respectively). These deviations may have multiple sources. The fundamental gaps of both materials could be underestimated, as we have mentioned above
for CsCu$_2$Br$_3$. Furthermore, the exciton binding energies are likely overestimated by $G_0W_0$+BSE, which is also observed for the strongly bound excitons in LiF \cite{Rohlfing98,Benedict98,Jiuyu2019}. To address this issue, one could apply self-consistent $G_0W_0$ with vertex correction, as well as an all-electron implementation;
this is, however, beyond the scope of this work.

\subsection{Performance of the Hybrid Functionals}\label{sec:IIIC}

\begin{figure}
	\centering
\includegraphics[width=8.0 cm]{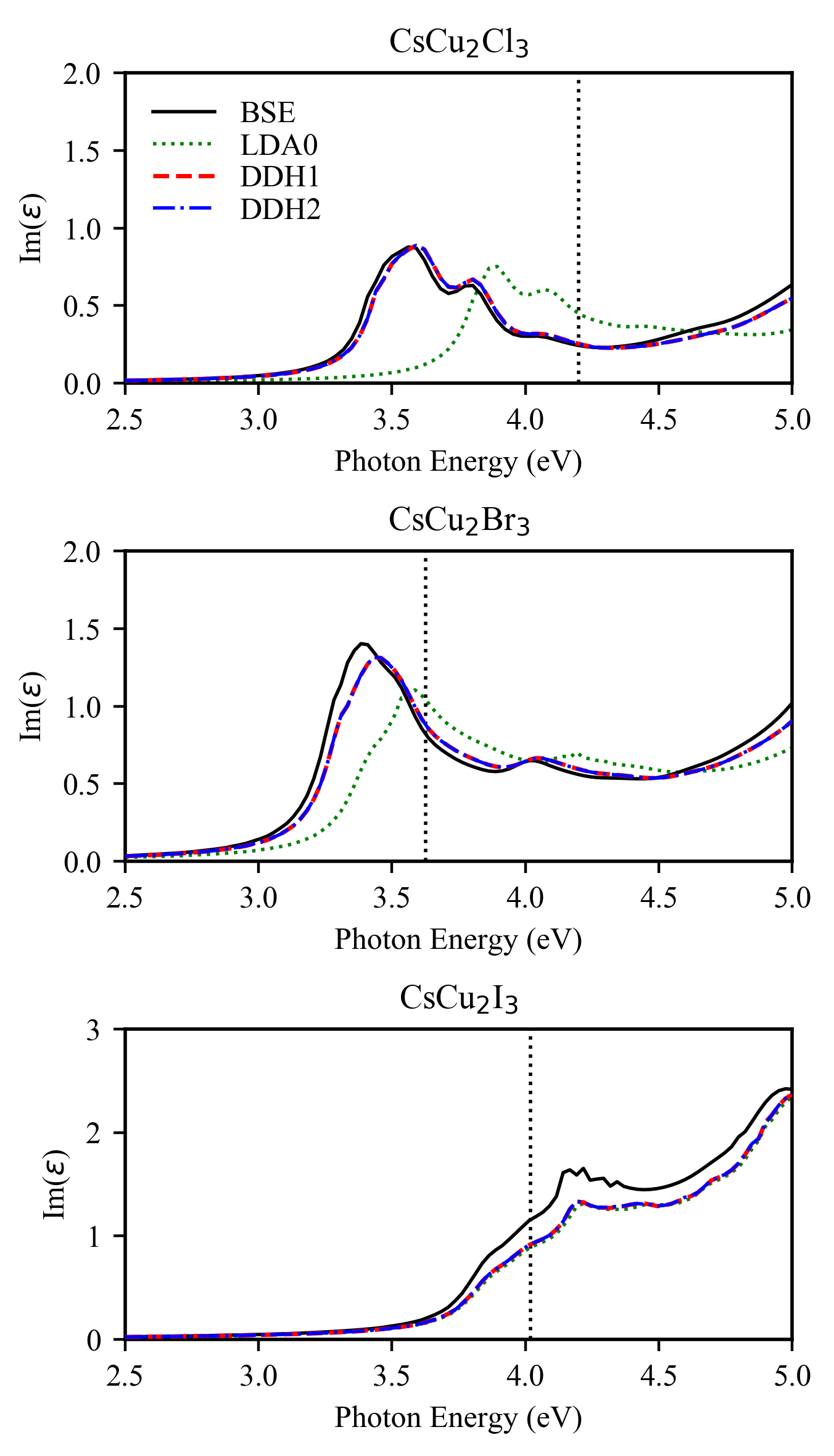}
     \caption{Optical spectra Im($\varepsilon$) along [111], calculated $G_0W_0$+BSE and hybrid functionals. The vertical dotted lines indicate
     the positions of the $G_0W_0$@LDA band gaps.}
     \label{fig:hyb_spc}
\end{figure}

We now compare the optical properties obtained by different hybrid functionals with BSE. Rather than discussing the [100], [010], and [001] spectra as in Section \ref{sec:IIIB},
we here only consider the spectra along [111], which can be viewed as averaging over the crystalline anisotropy. All calculated binding energies of the lowest energy excitons
are listed in Table~\ref{tbl:hyb}, and the optical spectra are plotted in Fig.~\ref{fig:hyb_spc}.

   \begin{table}
   \small
     \caption{Excitonic binding energies ($E_{\rm b}$, in meV) along [111], calculated using BSE and hybrid functionals, and RPA inverse dielectric constants
     ($\varepsilon^{-1}_{0,0}$).}
     \label{tbl:hyb}
     \begin{tabular*}{0.48\textwidth}{@{\extracolsep{\fill}}lccccc} 
     \hline
				&	$\varepsilon^{-1}_{0,0}$ & \multicolumn{4}{c}{$E_{\rm b}$ }  \\
					  	& 		& 	BSE  & LDA0 & DDH1 & DDH2\\\hline
   CsCu$_2$Cl$_3$   & 0.343 &  701 	&  423  &  682	   &  682  \\
   CsCu$_2$Br$_3$   & 0.298 & 321	  &  227  &  316 	  &	 317 \\
   CsCu$_2$I$_3$     & 0.254 &  156	   & 150   &  151     &	151	 \\\hline
     \end{tabular*}
   \end{table}

We first point out that the BSE $E_{\rm b}$ along [111] in Table~\ref{tbl:hyb} are close to or the same as the largest $E_{\rm b}$ in Table~\ref{tbl:optic} for each material. The slight enhancement of $E_{\rm b}$ for CsCu$_2$Cl$_3$ is because $\varepsilon^{-1}_{0,0}$ is larger along [111] than along [001]. The $\varepsilon^{-1}_{0,0}$ along [111] is in fact close to or the same as $\varepsilon^{-1}_{0,0}$ along [100] for each material, which indicates a dominant role played by the dielectric constant along [100] in these anisotropic materials. We then use the $\varepsilon^{-1}_{0,0}$  in Table~\ref{tbl:hyb} as the prefactors of Fock exchange in the DDH1 and DDH2 functionals.

For all three materials, LDA0 underestimates $E_{\rm b}$ compared to BSE. The LDA0 produces an error as large as 40\% for CsCu$_2$Cl$_3$. This can be clearly seen in the optical spectra in Fig.~\ref{fig:hyb_spc}, where the LDA0 spectrum is significantly blueshifted for CsCu$_2$Cl$_3$. The reason for the substantial LDA0 errors is the improper $a=0.25$, which leads to an overscreened Coulomb interaction. The error caused by the empirical $a$ becomes less severe going from CsCu$_2$Cl$_3$ to CsCu$_2$I$_3$. As  $\varepsilon^{-1}_{0,0}$ comes closer to $a=0.25$, the LDA0 error of $E_{\rm b}$ decreases to 29\% for CsCu$_2$Br$_3$ and to 3\% for CsCu$_2$I$_3$.

For the hybrid functionals DDH1 and DDH1 we have $a=\varepsilon^{-1}_{0,0}$, and $E_{\rm b}$ deviates less than 3.5\% from BSE. As shown in Fig.~\ref{fig:hyb_spc}, the
DDH spectra of CsCu$_2$Cl$_3$ are almost identical to the BSE spectra, apart from a tiny blueshift. For CsCu$_2$Br$_3$, the DDHs reproduce the slight shoulder
at about 3.3 eV. However, both DDHs underestimate the height of the dominant peak at about 3.4 eV, and shift it to about 3.5 eV. Comparing with the spectra in
Fig.~\ref{fig:bse_spc}, we conclude that this peak arises from the excitonic effects along [001]. Similar DDH results are found for CsCu$_2$I$_3$: $E_{\rm b}$
of the lowest energy exciton is very well reproduced, but the strength of the excitonic effects along [001] is somewhat underestimated. This suggests that the local-field
effects of the excitons in CsCu$_2$X$_3$ may not be fully captured by DDH1 and DDH2. We attribute this to the global prefactor of the Fock exchange;
using range-separated DDH functionals might improve the description. Nevertheless, it is evident that the nonempirical global DDHs are able to predict $E_{\rm b}$
and the optical spectra of CsCu$_2$X$_3$ with very good accuracy (using BSE as reference).

Finally, we point out the very close agreement between DDH1 and DDH2. From  Eqs. (\ref{K_hybrid1}) and (\ref{K_hybrid2}), the difference between DDH1 and DDH2 is $aE_{\rm c}$. For CsCu$_2$X$_3$, $a$ is in the range of 0.25-0.35. In weakly correlated materials such as those considered here, correlation effects are typically an order of magnitude
smaller than exchange effects. Thus, $aE_{\rm c}$ does not make much of a difference for the optical properties of CsCu$_2$X$_3$. However, we emphasize that there is no guarantee that the two functionals DDH1 and DDH2 perform the same way in other systems, especially materials with small dielectric constant (larger $a$) or stronger correlation.
In such cases, it might turn out that other forms of hybrid functionals, perhaps involving additional parameters, are preferable.

\section{Conclusions} \label{sec:IV}
In this work, we have shown that hybrid TDDFT is a very promising alternative to the BSE for calculating the optical properties of
semiconducting and insulating materials. To demonstrate this, we chose a class of perovskite materials, the Cs-Cu halides, with a
relatively complex structure containing 24 atoms per unit cell. These materials have a quasi-1D structure and strong excitonic features, giving rise to
light emission properties with potential for technological application.

Calculating the optical absorption spectrum of a material is a two-step process. First, the equilibrium band structure must be obtained, and
then the spectrum is calculated using linear response. In the many-body approach, the QP band structure is calculated using $GW$, followed by
the BSE. TDDFT instead takes Kohn-Sham band structures as input. Generalized TDDFT, using hybrid functionals, can be viewed as a bridge
between these two approaches.

Strictly speaking, the consistent way to implement generalized TDDFT is to calculate the band structure and the spectrum using the same hybrid
functional. In this paper, we instead chose to use the same $G_0W_0$ band structures as input for both TDDFT and the BSE, because our focus was not on
the performance of hybrid functionals for the band structure (which has been widely studied in the literature), but rather on the performance for
calculating optical spectra with prominent excitonic effects. Using different input band structures would have made a direct comparison difficult.
Furthermore, as shown in the Supplemental Material \cite{supp}, using hybrid band structures together with hybrid TDDFT does not lead to better optical
gaps than $G_0W_0$+BSE.

Here, we considered two different types of hybrid functionals: LDA0, which has a fixed, material-independent mixing parameter, and DDH1 and DDH2, with
a mixing parameter dependent on the dielectric constant of the material. The main finding is that only the dielectric-dependent hybrids can properly
account for the screening effects, and they do so extremely well: both DDH1 and DDH2 (which slightly differ in the way in which local exchange-correlation
is mixed with exact exchange) produce optical spectra that are very close to the BSE.

The computational cost of hybrid TDDFT can be significantly lower than that of the BSE; in a previous study \cite{Jiuyu2019}, substantial CPU time reductions
were achieved, depending on details of the implementation. Since those cost savings come at practically no loss of accuracy as we have seen here,
this clearly shows that hybrid TDDFT should become the method of choice for calculating excited-state properties of complex materials.

Dielectric-dependent hybrid functionals clearly warrant further study. Tests for additional classes of materials, including low-dimensional
structures, are necessary to find the optimal form of hybrid functionals for extended solids. In particular, a systematic comparison of dielectric-dependent
hybrids with and without range separation would be of interest. Another intriguing question is how dynamical screening effects, which are neglected in most
applications of the BSE \cite{Romaniello2009,Blase2018}, would carry over into the framework of generalized TDDFT.

\acknowledgments{This work was supported by NSF grant No. DMR-1810922. The computation for this work was performed on the high performance computing infrastructure provided by Research Computing Support Services at the University of Missouri-Columbia. }

\bibliography{CsCuX_paper}

\end{document}